\documentclass[aps,prd,preprint,superscriptaddress,showpacs]{revtex4}
\usepackage{epsf,epsfig,graphics,graphicx}
\usepackage{verbatim,color,ulem}
\newcommand{\be}{\begin{equation}}
\newcommand{\ee}{\end{equation}}
\newcommand{\bea}{\begin{eqnarray}}
\newcommand{\eea}{\end{eqnarray}}
\newcommand{\ba}{\begin{array}}
\newcommand{\ea}{\end{array}}
\usepackage{amssymb,amsmath,amsthm,graphicx}
\usepackage[mathscr]{eucal}
\usepackage{enumerate,color,verbatim,multirow,comment}

\begin{document}
\title{Subvacuum effects in Quantum Critical Theories from Holographic Approach}
\author{Chen-Pin Yeh}
\email{chenpinyeh@mail.ndhu.edu.tw} \affiliation{Department of
Physics, National Dong-Hwa University, Hualien, Taiwan, R.O.C.}

\author{Da-Shin Lee}
\email{dslee@mail.ndhu.edu.tw} \affiliation{Department of Physics,
National Dong-Hwa University, Hualien, Taiwan, R.O.C.}

\begin{abstract}
The subvacuum phenomena, induced by the squeezed vacuum of the
strongly coupled quantum critical fields with a dynamical scaling
$z$, are explored by a probe particle. The holographic description
corresponds to a string moving in 4+1-dimensional Lifshitz
geometry with gravitational wave perturbations.
The dynamics of the particle can be realized from the motion of
the endpoint of the string at the boundary. We then examine the
particle's velocity dispersion, influenced by the squeezed vacuum
states of the strongly coupled quantum critical fields. With
appropriate choices of squeezing parameters, the velocity
dispersion is found smaller than the value caused by the normal
vacuum fluctuations of the fields. This leads to the subvacuum
effect. We find that the large coupling constant of the quantum
fields tends to counteract the effect in reduction of velocity
dispersion, though this phenomenon is in principle observable. The
effect of the squeezed vacuum on the decoherence dynamics of a
quantum particle is also investigated. Coherence loss can be shown
less severe in certain squeezed vacuums than in normal vacuum.
This recovery of coherence is understood as recoherence, another
manifestation of the subvacuum phenomena. We make some estimates
of the degree of recoherence and find that, on contrary to the
velocity dispersion case,  the recoherence effect is enhanced by
the large coupling constant. Finally we compare the findings in
our earlier works when the particle is influenced by a weakly
coupled relativistic field with the dynamical scaling $z=1$.
\end{abstract}


\maketitle
\section{Introduction}
Engineering vacuum state may render suppression of its quantum
fluctuations, resulting in the so-called subvacuum phenomenon. The
existence of negative energy density of the quantum field is a
renowned example, where the renormalized expectation value of the
energy density operator can become negative in some spacetime
regions~\cite{EP}. However if the dynamics of quantum field theory
places no restrictions on negative energy density,  it may produce
significant macroscopic effects that potentially violate the
second law of thermodynamics~\cite{FO,DA} or the cosmic
censorship~\cite{FOR}. The negative energy density may also imply
exotic phenomena such as traversable wormholes~\cite{MO} and warp
drive~\cite{AL}. Thus, it has been shown that the renormalized
local energy density can not be arbitrarily negative for an
arbitrarily long period of time. There exists an inequality,
constraining the magnitude  and duration of the negative energy
density~~\cite{FOR,FO1,PF,FE}. Different aspects of the subvacuum
effects of quantum field theory can be realized by considering
these effects on the dynamics of a particle, with which the field
couples. It is found that the velocity dispersion of the particle,
induced by the squeezed vacuum of the quantum fields with
appropriate choices of squeezing parameters,  can be smaller than
the value determined by the normal vacuum~\cite{Lee_12}. A similar
subvacuum phenomenon has been investigated in the context of
quantum decoherence of the particle state as a result of an
unavoidable interaction with  environmental quantum
fields~\cite{HS,Lee_06}. This decoherence effect can be observed
by the contrast change of the fringes in the interference pattern.
It has been shown that coherence loss can be less severe in
certain squeezed vacuums than in normal vacuum. This is known as
recoherence. However, all of the above-mentioned subvacuum
phenomena are explored in either free or weakly coupled field
theories. Accordingly, their effects are supposedly rather weak.
This prompts us to pursue these subvacuum effects arising from
strongly coupled fields.

The idea of holographic duality was originally proposed as the
correspondence between 4-dimensional conformal field theory (CFT)
and  gravity theory in 5-dimensional anti-de Sitter (AdS)
space~\cite{AdSCFT}. It is soon generalized to other backgrounds
and field theories and opens up the possibility to study the
strong coupling problems in the condensed matter systems and the
hydrodynamics of the quark-gluon plasma
(see~\cite{Hartnoll_09,Rangmanai_09} for reviews). There have been
considerable efforts of employing the holographic duality to
explore the dissipative dynamics of a particle moving in a
strongly coupled
environment~\cite{Herzog:2006gh,Gubser_06,Teaney_06}. In these
cases, the endpoint of the string on the boundary of the AdS black
hole serves as a probed particle. Thereafter more works were
devoted to understanding the fluctuations of this endpoint of the
string in terms of Brownian motion in general
backgrounds~\cite{Son:2009vu,gubserqhat,ctqhat,Caceres:2010rm,Giecold:2009cg,CasalderreySolana:2009rm,Atmaja:2010uu,Das:2010yw,Gursoy:2010aa,Ebrahim:2010ra,CaronHuot:2011dr,Kiritsis:2011bw,
Fischler:2012ff,Tong_12,Edalati:2012tc,Atmaja:2012jg,Sadeghi:2013lka,Atmaja:2013gxa,Banerjee:2013rca,Giataganas:2013hwa,mirror,
Kiritsis:2013iba,Chakrabortty:2013kra,Sadeghi:2013jja,Giataganas:2013zaa,Sadeghi:2014lha,Fischler_14}.
A review on the holographic Brownian motion can be found in
\cite{Holographic QBM}.

The holographic duality provides a phenomenological description of
strongly coupled physics, but its applicability for a real
physical system has to be justified by its success in explaining
and predicting experimental results. Thus, our current work will
explore the subvacuum effects on a particle in squeezed vacuum of
strongly coupled quantum critical fields by the method of
holographic influence functional. This method was developed by
us~\cite{Yeh_14}, and is consistent with the one in
\cite{Son_09,Son_02,Ree_08}. It is hoped that the large coupling
constant might offer the possibility to observe these effects. The
dual description for a particle coupled to quantum critical fields
in its normal vacuum state corresponds to a string hanging from
the boundary of $4+1$-dimensional Lifshitz geometry. The endpoint
of a string is identified as the particle's position. Thus, the
effects of the quantum critical fields on the dynamics of the
particle will be encoded in the Green's functions of the
associated boundary fields. Additionally, a possible
holographically realization of the squeezing vacuum of the
boundary fields is given by the gravitation wave perturbations in
Lifshitz geometry~\cite{Lenny_99}. The corrections to the Green's
functions of the boundary fields can be found from the perturbed
holographic influence functional in bulk geometry, altered by the
gravitational wave. In the context of holographic Brownian motion,
the squeezed-vacuum correlation functions of the string's endpoint
can also be constructed through the Bogoliubov transformations
from the normal vacuum state. The corresponding Green's functions
of boundary fields can then be obtained via the Langevin equation,
derived from holographic influence functional.
Comparing these two results, we find the forms of the
Hadamard functions of boundary field, in leading order of the small
squeezing parameter and gravitational wave perturbations, allow
the identification of the squeezing parameter with the boundary
value of the gravitational wave perturbation. The Bogoliubov
transformations leave the retarded Green's function
unchanged. This can be confirmed by looking at the
holographic retarded Green's function to leading order in
gravitational wave perturbation. Then we are able to explore the
effects from squeezed vacuum of the strongly coupled fields on the
dynamics of a particle.

Our presentation is organized as follows. In next section, we briefly
review the method of holographic influence functional. The
environmental degrees of freedom in the full density matrix are
traced over to obtain the reduced density matrix of the system.
Their effects are all encoded in the influence functional. We then construct the
holographic influence functional for a
probed string in Lifshitz geometry. Later the
Lifshitz geometry is perturbed by gravitational waves.  The perturbed
holographic influence functional is found, from which  the nonequilibrium
Green's functions of boundary fields are obtained from this bulk construction. In Sec.~\ref{sec2}, the
correlation functions of the string's end point in its squeezed vacuum states can also be
constructed via the  Bogoliubov transformations from the normal
vacuum state of the string. With the derived Langevin equation of the string's end point, it allows to
identify the possible holographic dual of the squeezed vacuum states
as gravitation waves perturbations. We then study subvacuum
effects on particle's velocity dispersion influenced by squeezed
vacuum fluctuations. In Sec.~\ref{sec3}, the decoherence dynamics
of a particle affected by the squeezed vacuum of quantum critical
fields are also explored. The reduction in quantum coherence is
measured by the decoherence functional, given by the
holographic influence functional. We propose an interference
experiment to find the subvacuum effect of recoherence. Concluding
remarks are in Sec.~\ref{sec4}.

\section{Holographic influence functional and correlators for the squeeze vacuum states} \label{sec1}

 \subsection{Influence functional in field theory}
We first review the method of influence functional in field
theory. When the system of interest couples with the environment, their full dynamics can be described
by the density matrix ${ \rho}(t)$ that evolves unitarily
according to
\begin{equation}
{ \rho} (t_f) = U(t_f, t_i) \, { \rho} (t_i) \, U^{-1} (t_f,
t_i )
\end{equation}
with $ U(t_f,t_i) $ the time evolution operator of the system and
environment.  The reduced density matrix $\rho_r$ is obtained
by tracing over the environmental degrees of freedom in the full density matrix, and it will include all the
effects from the environment on the system. We assume that the
initial density matrix at time $t_i$ is factorized as
\begin{equation}\label{initialcond}
    \rho(t_i)=\rho_{q}(t_i)\otimes\rho_{{F}}(t_i)\,
    ,
\end{equation}
for simplicity, where $q$ and $F$ generically represent the system and the environment variables respectively.
We further assume that the environment field  is in thermal
equilibrium at temperature $T=1/\beta $ before it is brought into contact with the system, so
$\rho_{{F}}(t_i)$ takes the form
\begin{equation}\label{initialcondphi}
    \rho_{F}(t_i)=e^{-\beta H_{F}}\,,
\end{equation}
where $H_{F}$ is the Hamiltonian for the $F$ field. The
vacuum state  can be obtained by taking the zero-$T$ limit. We
consider the system linearly coupled to an environment field.
The full Lagrangian  takes this form
  \be
   L(q,F)=L_q[q]+L_F[F]+qF\, .
  \ee
Then the reduced density matrix  becomes~\cite{Leggett,SK,GSI}
\begin{equation}
\rho_r({q}_f,\tilde{{q}}_f,t_f)=\int\!d{q}_1\,d{q}_2\;\mathcal{J}({q}_f,\tilde{{q}}_f,t_f;{q}_1,{q}_2,t_i)\,\rho_{q}({q}_1,{q}_2,t_i)\,,\label{evolveelectron}
\end{equation}
where the propagating function
$\mathcal{J}({q}_f,\tilde{{q}}_f,t_f;{q}_1,{q}_2,t_i)$ is
\begin{equation}\label{propagator}
    \mathcal{J}({q}_f,\tilde{{q}}_f,t_f;{q}_1,{q}_2,t_i)=\int^{{q}_f}_{{q}_1}\!\!\mathcal{D}{q}^+\!\!\int^{\tilde{{q}}_f}_{{q}_2}\!\!\mathcal{D}{q}^-\;\exp\left[i\int_{t_i}^{t_f}dt\left(L_{q}[{q}^+]-L_{q}[{q}^-]\right)\right]\mathcal{F}[{q}^+,{q}^-]\,.
\end{equation}
The influence functional $\mathcal{F}[{q}^+,{q}^-]$ can be written
in terms of the real-time Green's functions~\cite{Fv},
\bea\label{influencefun2}
{\mathcal{F}}\left[{q}^{+},{q}^{-}\right]&& = \exp\bigg\{
-\frac{i}{2}\int_{t_i}^{t_f} dt\!\!\int_{t_i}^{t_f} \!dt' \Big[
{q}^+(t)\,G^{++}(t,t') \,{q}^+(t')\Bigr.\bigr.
-{q}^+(t)\,G^{+-}(t,t')\,{q}^-(t') \nonumber\\
&&- {q}^-(t)\,G^{-+}(t,t')\,{q}^+(t')
\big.\big.+{q}^-(t)\,G^{--}(t,t')\,{q}^-(t')\big]\bigg\}\,,
 \eea
 where we keep the terms to the quadratic order in the particle position, valid in the linear response approximation.
 The Green's functions involved are time-ordered, anti-time-ordered and Wightman functions, defined
as  \bea \label{correlator}
    && i\,G^{+-}(t,t')=\langle F(t')F(t)\rangle \, , \nonumber\\
    && i\,G^{-+}(t,t')=\langle F(t)F(t')\rangle \, ,\nonumber\\
    &&i\,G^{++}(t,t')=\langle F(t)F(t')\rangle\theta(t-t')+\langle
    F(t')F(t)\rangle\theta(t'-t)\, , \nonumber\\
    && i\,G^{--}(t,t')=\langle F(t')F(t)\rangle\theta(t-t')+\langle
    F(t)F(t')\rangle\theta(t'-t) \, .
    \eea
The retarded Green's function and Hadamard function can be
constructed from them according to  \bea \label{G_HR}
    G_{R} (t-t')&\equiv & -i \theta (t-t')\langle [F(t), F(t')] \rangle = \bigg\{ G^{++} ( t,t') -G^{+-} (t,t') \bigg\}\,,  \\
    G_{H} (t-t')&\equiv & \frac{1}{2} \langle \{ F(t), F(t') \} \rangle=\frac{i}{4} \bigg\{ G^{++} ( t,t') +G^{+-} (t,t') + G^{--} ( t,t') +G^{-+} (t,t')
    \bigg\}\,. \nonumber
\eea In a time-translation invariant environment, the  Fourier transform
of various Green's functions is defined by
\begin{equation} G (t-t')=\int \, \frac{ d \omega}{2\pi} \, G
(\omega) \, e^{-i \omega (t-t')} \, . \end{equation} Their
respective $\omega$-dependent functions are obtained as
  \bea
  \label{SKs}
  G^{++}(\omega)&=&{{\rm Re}}G_R(\omega)+(1+2n)\,i\,{\rm Im}G_R(\omega)\, ,\nonumber\\
  G^{--}(\omega)&=&-{\rm Re}G_R(\omega)+(1+2n)\,i\,{\rm Im}G_R(\omega)\, , \nonumber\\
  G^{+-}(\omega)&=& 2n\,i\,{\rm Im}G_R(\omega)\, ,\nonumber\\
  G^{-+}(\omega)&=& 2(1+n)\,i\,{\rm Im}G_R(\omega)
  \,
  \eea
with $n=(e^{\frac{\omega}{T}}-1)^{-1}$.
Notice that the above Green's functions are not totally independent
and they obey the following relations:
\begin{equation}
\qquad\qquad\qquad G^{++}(\omega) +  G^{--}(\omega)- G^{+-}(\omega)-
G^{-+}(\omega) =0 \, , \label{u-identity}
\end{equation}
 resulting from the unitary evolution of the system and environment, and
  \be \label{periodical_T}
  \frac{G^{+-}(\omega)}{G^{-+}(\omega)}=e^{-\frac{\omega}{T}}
  \,
  \ee
due to a bosonic thermal bath. Additionally  the fluctuation-dissipation
relation gives \be \label{FD} G_H(\omega)=- (1+2n) \, {\rm Im} G_R
(\omega) \, .\ee

\subsection{Holographic influence functional}
In this section we review the construction of the holographic
influence functional from dual gravity theory. The conventional
approach to derive the influence functional can at best be
perturbatively implemented for a weakly coupled environment, let
alone the strong coupled theory. Thus for the latter case,  the
holography method will be employed to find the influence
functional of the strongly coupled environment. We consider a
particle coupled to quantum critical theories in 3+1-dimension at
zero temperature~\cite{Tong_12,mirror}. Its dual description is a
straight string moving in the 4+1-dimensional Lifshitz geometry
with the metric
  \be
  \label{bmetric}
  ds^2=-\frac{r^{2z}}{L^{2z}} dt^2+\frac{L^2}{r^2} dr^2 +\frac{r^2}{L^2} d\vec{x}^2 \, ,
  \ee
in which $L$ is the radius of curvature of the geometry. This
gravity background~(\ref{bmetric}) can be engineered by coupling
the gravitation field with negative cosmological constant to a
massive vector field. The relevant action to give the above metric
is \cite{Taylor},
  \be
  \label{action}
  S=\frac1{16\pi G_{4+1}}\int
  d^{4+1}x \,\sqrt{-g}\, ( R+ 2\Lambda-\frac14 {\cal F}^{\mu\nu}{\cal F}_{\mu\nu}-\frac1 2m_A^2 {\cal A}^{\mu} {\cal
  A}_{\mu}) \, .
  \ee
In addition to the Einstein-Hilbert action and the cosmological
constant $\Lambda$ term, the action of a vector field ${\cal
A}_{\mu}$ with mass $m_A$ is introduced where ${\cal F}_{\mu\nu}$ is
the field strength of ${\cal A}_{\mu}$. The equations of motion
for the metric and the vector field can be derived from this
action, and they are
   \bea \label{EoM}
   &&R_{\mu\nu}=-\frac{2}{3}\Lambda g_{\mu\nu}+\frac12 g^{\alpha\beta}{\cal F}_{\mu\alpha} {\cal F}_{\nu\beta}+\frac12 m_A^2 {\cal A}_{\mu} {\cal A}_{\nu}-\frac1{12} {\cal F}_{\alpha\beta} {\cal F}^{\beta\alpha}g_{\mu\nu}\, ,\\
   &&D_{\mu}{\cal F}^{\mu\nu}=m_A^2 {\cal A}^{\nu} \, ,
   \eea
where $D_{\mu}$ is a covariant derivative with respect to the
background metric $g_{\mu \nu}$. The solutions of the vector field
are assumed to be
  \be \label{A_field}
  {\cal A}^{(0)}_{\mu }={\cal A} \frac{r^z}{L^z}\delta^{0}_{\mu} \, .
  \ee
Then the Lifshitz background with metric~(\ref{bmetric}) can be
achieved by setting
 \be
 \label{para}
 {\cal
 A}=\sqrt{\frac{2(z-1)}{z}},~~m_A^2=\frac{3 z}{L^2},~~\Lambda=\frac{9 +2 z-z^2}{2L^2}
 \, .
 \ee

Using the method of holography, as in \cite{Tong_12,mirror}, the
classical on-shell gravity action of the string moving in the
Lifshitz background can be identified as the influence functional
for a particle in strongly coupled field theory. We start from
considering the finite-temperature nonequilibrium correlators and
the zero temperature limit will be taken later. Hereafter, $L=1$
will be used, and reintroduced later when needed. The metric of
Lifshitz black hole has the form
  \be
  \label{lifshitz bh}
  ds^2=-r^{2z}f(r)dt^2+\frac{dr^2}{f(r)r^2}+r^2d\vec{x}^2 \, ,
  \ee
where $f(r)\rightarrow1$ for $r\rightarrow\infty$ and $f(r)\simeq
c(r-r_h)$ near the black hole horizon $r_h$ with
$c=({z+3})/{r_h}$. The detailed form of $f(r)$ is not relevant
since only the low-frequency  perturbation is
considered in our subsequent discussions. The black hole temperature, which is also the temperature
in the boundary field theory, is
 \be
 \label{BHT}
\frac1T=\frac{4\pi}{z+3}\frac1{r_h^z} \, .
  \ee
Here we assume that the string moves only along the $x$-direction
and its position variable is $X$. It is then  straightforward to
obtain the linearized Nambu-Goto action for a string in the
background of  Lifshitz black hole: \be
  S_{NG} =- \frac1{4\pi\alpha'}\int dr \, dt \,
\bigg( r^{z+3} \, f(r)  \,  \partial_rX
\partial_rX-  \, \frac{\partial_tX\partial_tX}{f(r){r^{z-1}}}\bigg)
\, , \label{S_NG_0}
  \ee
where $X(t,r)$ is the   linearized string perturbation in the static
gauge. The equation of motion of the string is
 \be
  \label{NG with T}
 \frac{\partial}{\partial r}\biggl[ r^{z+3}\, f(r) \, \frac{\partial }{\partial
 r} X (r, t) \biggr]-\frac{\partial}{\partial t}\, \biggl[ \frac{1}{r^{z-1}}\frac1{f(r)} \,
 \frac{\partial}{\partial t} X (r, t)\biggr]=0
 \, .
  \ee
We express the solutions in terms of two linearly
independent solutions, {whose Fourier transformations are complex conjugates to one another, $\mathcal{X}_{\omega}(r)$ and $\mathcal{X}^{*}_{\omega} (r)$
and have the properties $\mathcal{X}_{\omega}(r)_{\substack{
   \propto \\
   r\rightarrow r_h
  }} e^{+i\omega r_*}$ and
$\mathcal{X}^{*}_{\omega}(r)_{\substack{
   \propto \\
   r\rightarrow r_h
  }} e^{-i\omega r_*}$.
Here we have defined
  \be
  r_*=\int dr\frac1{r^{z+1}f(r)} \, ,
  \ee}
and the normalization is such that $\mathcal{X}_{\omega}(r_b)=1$.

In accordance with the
closed-time-path formalism~\cite{Leggett,SK,GSI} we have discussed in the previous section, we introduce
$Q^+(t,r_1)$ and $Q^-(t,r_2)$, which are the string worldsheets
living  in
maximally extended black hole geometry with two asymptotical  boundaries~\cite{Son_09, Son_02}. Then, by choosing appropriate boundary conditions for the
perturbations of the string $Q^{\pm}(t, r)$ in this background geometry,
the classical on-shell action of the string can be identified as
the influence functional for a particle affected by the boundary fields~\cite{Son_09}:
   \be
   \label{gravity action}
   \mathcal{F} [q^+,
   q^-]=S_{gravity}\left(Q^+(t,r_b),Q^-(t,r_b)\right)\, ,
   \ee
   where the gravity action is the action $S_{NG}$
   in~(\ref{S_NG_0}).
After some algebraic reduction, the on-shell action
contains only the boundary terms and it takes the form
  \be \label{S_NG_onshell} S_{NG}^{{\rm on-shell}} \simeq
-\frac{r_b^{(z+3)}}{4 \pi \alpha'} \int dt \, \left( Q^+ (t,r_b)
\partial_r Q^+ (t,r_b)-  Q^- (t,r_b) \partial_r Q^- (t, r_b)
\right)\, .
  \ee
The parameter $r_b$ is the location of the
boundary and serves as a cutoff scale in the radial direction to
render the action finite. {The boundary conditions of the string
perturbations  are
  \be
 \label{bc} q^{\pm}(t)=Q^{\pm}(t,r_{b})\,,
 \ee
in which the variable $q (t)$ can be identified as the position of the
 Brownian particle.} Following~\cite{Yeh_14},
 which is  consistent with~\cite{Son_09, Son_02}, we find  $Q^{\pm} (\omega,r)$ given by
    \bea \label{Q_pm}
    &&Q^+(\omega,r_1)=\frac{1}{1-e^{-\frac{\omega}{T}}} \bigg[ (q^-(\omega)- e^{-\frac{\omega}{T}}q^+(\omega))
    \mathcal{X}_{\omega}(r_1)+ ( q^+(\omega)-q^-(\omega))
 \mathcal{X}_{\omega}^*(r_1) \bigg]\, ,\nonumber\\
    &&Q^-(\omega,r_2)= \frac{1}{1-e^{-\frac{\omega}{T}}} \bigg[ (q^-(\omega)- e^{-\frac{\omega}{T}}q^+(\omega))
    \mathcal{X}_{\omega}(r_1)+ e^{-\frac{\omega}{T}} ( q^+(\omega)-q^-(\omega))
 \mathcal{X}_{\omega}^*(r_1) \bigg] \, .  \eea
This general solution is then substituted into the classical
on-shell action~(\ref{S_NG_onshell}). Using (\ref{influencefun2})
and (\ref{G_HR}), the retarded Green's function at finite
temperature is obtained to be
   \be
   \label{G_RT}
{G_R(\omega)= \frac{r_b^{z+3}}{2\pi\alpha'}
\mathcal{X}_{-\omega}(r_b)\partial_r\mathcal{X}_{\omega}(r_b)}\,.
  \ee
In general, the analytical expression of $G_R(\omega)$ is not available
except in the small $\omega$ limit. Nevertheless, there is an analytical
solution for $\mathcal{X}_{\omega}(r)$ at zero temperature,
  \be
  \label{mode}
  \mathcal{X}_{\omega}(r)=\frac{r_b^{1+\frac{z}2}}{r^{1+\frac{z}2}}\frac{H^{(1)}_{\frac1z+\frac12}(\frac{\omega}{zr^z})}{H^{(1)}_{\frac1z+\frac12}(\frac{\omega}{zr_b^z})}
  \, .
  \ee
Hence the zero-temperature retarded Green's function can be found
for $\omega
>0$ to be~\cite{Tong_12},
  \be
\label{G_R} G^{(0)}_R(\omega)=- \frac{\omega
r_b^2}{2\pi\alpha'}\frac{H^{(1)}_{\frac1z-\frac12}(\frac{\omega}{zr_b^z})}{H^{(1)}_{\frac1z+\frac12}(\frac{\omega}{zr_b^z})}
\,.
  \ee
All other correlators can be derived from~(\ref{SKs}) by taking the
$T\rightarrow 0$ limit. In particular, through the
fluctuation-dissipation relation~(\ref{FD}) in the $T \rightarrow 0$ limit, we obtain the Hadamard
function for $\omega
>0$ as follows:
 \be \label{G_0_H}
 G^{(0)}_H(\omega)=\frac{z r_b^{2+z}}{\pi^2\alpha'}\frac1{J^2_{\frac1z+\frac12}(\frac{\omega}{zr_b^z})+Y^2_{\frac1z+\frac12}(\frac{\omega}{zr_b^z})}
  \, . \ee

\subsection{Influence functional from gravitational wave perturbed geometry}
We can engineer the vacuum of the boundary fields by  perturbing the bulk
geometry. In particular, as suggested in \cite{Lenny_99}, we
consider the gravitational wave perturbations in the Lifshitz background
with the metric $g_{\mu\nu}^{(0)}$~(\ref{bmetric}). The perturbed
metric is given by
   \be
   \label{gravitonp}
   g_{\mu\nu}=g^{(0)}_{\mu\nu}+r^2\phi(t,r)\xi_{\mu\nu} \, ,
   \ee
where $\xi_{\mu\nu}$, the polarization tensor, has non-zero
components only in the spatial directions of the boundary, and is
assumed transverse and traceless. The field $\phi (t,r)$ is
introduced to parameterize small metric perturbations from
gravitation waves, and its the equation of motion is,
   \be
   \label{eom}
 {r^{-2z}}\, \partial_t^2\phi(t,r)+(3+z)\, r\partial_r\phi
(t,r)+r^2\,
   \partial_r^2\phi(t,r)=0 \,
   \ee
which is obtained by linearizing~(\ref{EoM}) about the background
solutions~(\ref{bmetric})  and~({\ref{A_field}).  Thus, the Fourier transform of the $\phi(t,r)$ field in frequency space is defined  as
\be
  \label{phi}
  \phi (t,r)= \int_{0}^{\infty}
  d\omega\, \phi (\omega,r)   \, e^{-i \omega t}+{\rm h. c.}.
  \ee
The normalizable solution of  (\ref{eom}) can be given by
\be
  \label{varphi}
 \phi (\omega,r)  =r^{-\frac{2+z}{2}} \,  \varphi (\omega) \, J_{\frac{2+z}{2z}}\bigg(\frac{ \omega}{zr^z}\bigg)  \, .
  \ee
The function $\varphi (\omega)$ is determined by the boundary
condition of the gravitation waves at $r=r_b$, and will be
identified as the squeezing parameter defined below.  Here we also
assume that the string moves only along the $x$-direction with the
position variable $X$. Then the Nambu-Goto action in perturbed
Lifshitz geometry can be written explicitly as \be
  S_{NG} =- \frac1{4\pi\alpha'}\int dr \, dt \,
\bigg( r^{z+3} \, \,(1+ \phi (r, t)) \,  \partial_rX
\partial_r X -  \,(1+ \phi (r, t)) \, \frac{\partial_tX\partial_tX}{{r^{z-1}}}\bigg)
\, . \label{S_NG}
  \ee
 Up to the first order in $\phi$, the equation of motion of the string
becomes
 \be
  \label{NG with Ts}
 \frac{\partial}{\partial r}\biggl[ r^{z+3}\,  \,(1+ \phi (r, t))\, \frac{\partial }{\partial
 r} X (r, t) \biggr]-\frac{\partial}{\partial t}\, \biggl[ \frac{1}{r^{z-1}}  \,(1+ \phi (r, t))\,
 \frac{\partial}{\partial t} X (r, t)\biggr]=0
 \, .
  \ee
We consider the perturbative solution which in frequency space is
given by
  \be
   X_{\omega}(r)=X^{(0)}_{\omega}(r)+X^{(\phi)}_{\omega}(r) \, ,
  \ee
where the zeroth order solution~(\ref{mode}) gives the retarded
Green's function and Hadamard function in the unperturbed Lifshitz
spacetime. Then the equation of motion for
$X^{(\phi)}_{\omega}(r)$ to leading order is given by
  \bea
  \label{PertuEoM}
  \frac{\partial}{\partial r}
  \bigg[ r^{z+3}\partial_rX^{(\phi)}_{\omega}(r)\bigg]+ r^{1-z}\omega^2X^{(\phi)}_{\omega}(r)&&=- \int
  d\omega'\,  \bigg[ r^{z+3}\, \partial_r\phi (\omega, r)\, \partial_r
  X^{(0)}_{\omega-\omega'}(r) \nonumber\\
  && \quad\quad\quad\quad +\omega'(\omega-\omega')\, r^{1-z}\phi (\omega', r)\, X^{(0)}_{\omega-\omega'}(r)
  \bigg] \, .
  \eea
 In the small
$\frac{\omega}{r^z}$ limit, the asymptotical form of the
inhomogeneous solution is
  \be \label{x_phi}
 X^{(\phi)}_{\omega}(r)=\frac{r^{-2-3z}}{2z(2+3z)} \, \int
d\omega' \,
 \omega'(\omega-\omega')\bigg(\frac{\omega'}{z}\bigg)^{\frac{2+2z}{z}} \varphi (\omega')\,
  \ee
  with $\varphi (\omega)$ given by~(\ref{varphi}).
Thus, the corrections to the Nambu-Goto action due to the gravitation waves in~(\ref{S_NG}) have the explicit $\phi$ dependence and the contributions from $X^{(\phi)}_{\omega}(r)$.
Since $\phi (\omega, r_b) >> X_{\omega}^{(\phi)} (r_b)/
X_{\omega}^{(0)} (r_b)$ for large $r_b$,  the contributions from
$X_{\omega}^{(\phi)}$ to the above perturbed action \eqref{S_NG} can be ignored
if we keep the leading-order terms in small $\phi$. The on-shell perturbed
action $S^{{\rm on-shell}}_{NG \, \phi}$ is then expressed as
 \bea \label{S_phi} S^{\rm on-shell}_{NG \, \phi} &=& -\frac{r_b^{3+z}} { 4 \pi
\alpha'} \int dt \, \phi (t,r_b) \, \left( Q^+ (t,r_b)
\partial_r Q^+ (t,r_b)-  Q^-
(t,r_b) \partial_r Q^- (t, r_b) \right) \, \nonumber\\
&=& -\frac{r_b^{3+z}} { 4 \pi \alpha'} \int \frac{d\omega}{2\pi}
\int \frac{d \omega'}{2\pi}\, \phi(\omega+\omega',r_b)\left(
Q^+_{-\omega} (r_b) \,
\partial_r Q^+_{-\omega'} (r_b)- Q^-_{-\omega} (r_b) \,
\partial_r Q^-_{-\omega'} ( r_b)\right)\, .  \nonumber\\
 \eea
The holographic perturbed influence functional will be
found when we substitute into the
above expression the zero-$T$ limit of~(\ref{Q_pm}) with $ \mathcal{X}_{\omega}(r)$ given by (\ref{mode}). The corrections to the nonequilibrium Green's functions can be identified in this perturbed holographic influence functional. It will be shown later that
 the possible holographic description of squeezed vacuum states is given by the gravitation wave perturbations.
 Thus, these nonequilibrium Green's functions obtained from the perturbed influence functional will be compared with the Green's functions
of general multi-mode squeezed vacuum states of the boundary
fields. In so doing, the function $\varphi$ in~(\ref{varphi}),
determined by the boundary condition of gravitation waves, will be
related to the squeezing parameters of the squeezed vacuum states.

\section{velocity fluctuations of a particle} \label{sec2}
In principle the effects from vacuum fluctuations of an environment field on
 a particle can be revealed in the
particle's velocity dispersion. The environment
field  not only modifies the
evolution of the particle's mean trajectory but also introduces additional
stochastic motion~\cite{HWL}. These two effects are encoded in the
associated Langevin equation. To derive this equation from the
influence functional~(\ref{influencefun2}), we find it more convenient to change the $q^{+}$, $q^{-}$
coordinates to the average and relative coordinates:
\begin{equation}
 q= (  q^+ + q^- )/2 \, , \,\,\,\,\,
q_{\Delta}= q^+ - q^- \, .
\end{equation}
As will be seen later, the
influence of the environment field can give the mass of the particle, and
damp its motion. As such, here we consider that all terms associated with the particle are dynamically generated from the contributions of the environmental quantum fields.
Thus, the coarse-grained effective action can be defined from~(\ref{propagator}) with the influence functional~(\ref{influencefun2}) only, and thus there is no need to introduce the Lagrangian $L_q[q]$ of the particle: 
 \bea
 & & S_{\rm CG} \big[q^{\pm} = q \pm q_{\Delta}/2 \big]  = -i   \,  \ln{\cal
F} \left[  q^+,  q^- \right] \nonumber\\
&& \quad = \int dt  q_{\Delta}(t) \left[ - \int dt' G_R (t, t')
q(t') \right] + \frac{i}{2}  \int dt \int dt' q_{\Delta}(t) \, G_H
(t, t')  q_{\Delta} (t') \, .  \eea
 We then further introduce an
auxiliary variable $ \eta (t)$, the noise force,  with a Gaussian
distribution function:
\begin{equation}
P[\eta (t)] = \exp \left\{ - \frac{1}{2}  \, \int dt \, \int dt' \,
\eta (t) \, G^{-1}_{H} (t, t') \, \eta (t') \right\} \, .
\label{noisedistri}
\end{equation}
In terms of the noise force $\eta (t) $,  $ S_{CG} $ can be
rewritten as an ensemble average over $\eta (t)$,
\begin{equation}
\exp i S_{CG}  =  \int {\cal D} \eta \, P [\eta (t)] \, \exp i
S_{\eta} \left[ q, q_{\Delta} ; \eta \right] \, ,
\end{equation}
where  the stochastic coarse-grained effective action   $ S_{\eta }$
is given by \be S_{\eta} [q, q_{\Delta}  ; \eta ] = \int dt \,
q_{\Delta} (t) \left[- \int dt' \, G_R (t, t') \, q(t')+ \eta (t)
\right] \, . \ee Varying the action $S_{\eta}$ with respect to $q_{\Delta}$ and setting $q_{\Delta}=0$ will give the
Langevin equation of $(q(t)=X(t))$:
  \be
  \label{langevin_eq} \int dt' \, G_R (t, t') \, X(t')=\eta (t) \, .
  \ee
The noise force correlation function  can be obtained from
(\ref{noisedistri}) as
\begin{equation}
\langle \eta (t) \, \eta (t') \rangle =  \, G_H (t, t') \, . \label{langevin1}
\end{equation}
Evidently, the retarded Green's function will modify the trajectory, given by noise or external forces, and the correlated noise forces  will render the trajectory
fluctuating. Since the low frequency
expansion of the retarded Green's function at zero-$T$ in~(\ref{G_R})
is~\cite{Tong_12,mirror}
 \be \label{G_R_LowF}
  G_R^{(0)} (\omega)=m (z) (i\omega)^2+\mu (\omega,z) \, ,
  \ee
  for a general $z$ the induced mass $m$ and the $\mu$ term are given by
  \be \label{m_mu}
  m (z)=\frac{1}{\pi \alpha' (2-z)r_b^{z-2}},\qquad\qquad \mu (\omega,z)=\gamma(z)(-i\omega)^{1+\frac{2}z}+{\mathcal O}(\omega^4)
  \ee
with  the damping coefficient
 \be \label{gamma} \gamma (z)=\frac{1}{\pi \alpha'
(2z)^{2/z}}\frac{\Gamma(\frac12-\frac{1}{z})}{\Gamma(\frac12+\frac{1}{z})}
\,.
 \ee
Although both $m$ and $\gamma$ change their signs at $z = 2$, the
ratio $\gamma/m$ remains positive and varies continuously at
$z=2$. When $z>2$, the minus signs of the mass term and the term
associated with damping can be simultaneously removed in the
dynamical equation~(\ref{langevin_eq}) by changing the sign of the
noise forces, i.e. $\eta \rightarrow -\eta$, thus leading to a
sensible equation of motion.  In fact, the noise forces are
introduced as auxiliary variables and their effects on the
 dynamics of the particle are formulated only in the form of the correlation
functions. Then, the magnitude of $ m $ can be identified as a
dynamical mass  of the particle, which can be large, about the
order $\lambda \propto 1/ \alpha'$. $\lambda$ corresponds to the
coupling constant in quantum field theory via AdS/CFT
correspondence.

Before proceeding further, it is of interest to see how the
solution to the Langevin equation is connected with the
fluctuation-dissipation relation in~\cite{Tong_12}. In the case of
the vacuum state of an environment field, the Fourier transform of
the Langevin equation~(\ref{langevin_eq})  is expressed as
  \be
 X(\omega)
=\eta(\omega)/G_R^{(0)} (\omega) \, \label{Langevin_sol}.
  \ee
Then, the fluctuations in the position of a particle induced by
noise forces are obtained as
  \be \langle X(\omega) X(-\omega)
\rangle= \frac{\langle \eta(\omega) \eta(-\omega) \rangle}{G_R^{(0)}
(\omega) G^{(0) *}_R (\omega)}= \frac{G_H^{(0)} (\omega)}{G_R^{(0)} (\omega) G^{(0)*}_R
(\omega)}=- {\rm Im} \chi (\omega) \, ,
  \ee
where $\chi^{-1}(\omega)=-G_R^{(0)} (\omega)$ and the
fluctuation-dissipation relation~(\ref{FD}) in the $T \rightarrow 0$ limit is applied. The above expression is a key
result in~\cite{Tong_12}. Here we recover it by solving the
Langevin equation,  which is derived from  the obtained influence functional.

To proceed for the squeezed vacuum states, we consider a quantized
string, as in~\cite{Tong_12}, with its mode expansion as follows:
\begin{equation}\label{force_mode_exp}
    X(t)=X(t,r_b)=\sqrt{
    \frac{\pi \alpha'}{z}}\,\int_{0}^{\infty}\!d\omega\;U_{\omega} \, \Bigl[ a_{\omega}^{\vphantom{\dagger}}\,e^{-i\omega t}+a^{\dagger}_{\omega}\,e^{+i\omega t}\Bigr]\,,
\end{equation}
where  $a_{\omega}$ and $a_{\omega}^{\dagger}$ are the
annihilation and creation operators, and they obey  canonical
commutation relations. In the background of Lifshitz black hole,
the string perturbations are in thermal states  where $\langle
a^{\dagger}_\omega a_{\omega}\rangle=
(e^{\frac{\omega}{T}}-1)^{-1}$ with black hole temperature $T$.
The mode functions in zero-$T$ limit becomes
\begin{equation} \label{U}
   U_{\omega}=\frac{1}{\sqrt{1+C_{\omega}^{2}}}\frac{1}{r_b^{1+\frac{z}{2}}}\Bigl[J_{\frac{1}{2}+\frac{1}{z}}(\frac{\omega}{zr_b^{z}})+C_{\omega}\,Y_{\frac{1}{2}+\frac{1}{z}}(\frac{\omega}{zr_b^{z}})\Bigr]\,,
\end{equation}
in which
\begin{equation}
    C_{\omega}=-\frac{J_{-\frac{1}{2}+\frac{1}{z}}(\frac{\omega}{zr_b^{z}})}{Y_{-\frac{1}{2}+\frac{1}{z}}(\frac{\omega}{zr_b^{z}})}\,.
\end{equation}
Thus it is quite reasonable to assume that the squeezed vacuum states can
be constructed from the Bogoliubov transformations of the creation
and annihilation operators of the normal vacuum state. Here we
consider the two-mode squeezed states where the squeezed operator is
defined as
\begin{align}
    \lvert\xi_{\omega\omega'}\rangle&=S(\xi_{\omega\omega'})\,\lvert0\rangle\,,&S(\xi_{\omega\omega'})&=\exp\left[\frac{1}{2}\bigl(\xi_{\omega\omega'}^{*}\,a_{\omega}a_{\omega'}-\xi_{\omega\omega'}\,a_{\omega}^{\dagger}a_{\omega'}^{\dagger}\bigr)\right]\,
\end{align}
with the squeezing parameter
$\xi_{\omega\omega'}=r_{\omega\omega'}\,e^{i\theta_{\omega\omega'}}$.
With the help of the Baker-Campbell-Hausdorff formula, we readily
find the Bogoliubov transformations of the creation and
annihilation operators due to the squeeze operator
$\mathcal{S}\left(\xi_{\omega\omega'}\right)$,\begin{eqnarray}
    &&\mathcal{S}^{\dagger}\left(\xi_{\omega\omega'}\right)a_{\omega}\,\mathcal{S}\left(\xi_{\omega\omega'}\right)=\mu_{\omega\omega'}a_{\omega}-\nu_{\omega\omega'}a^{\dagger}_{\omega'}\,,\qquad\text{and}\qquad\mathcal{S}^{\dagger}\left(\xi_{\omega\omega'}\right)a^{\dagger}_{\omega}\mathcal{S}\left(\xi_{\omega\omega'}\right)=\mu_{\omega\omega'}a^{\dagger}_{\omega}-\nu^*_{\omega\omega'}a_{\omega'}\,,\nonumber\\
&&\mathcal{S}^{\dagger}\left(\xi_{\omega\omega'}\right)a_{\omega'}\,\mathcal{S}\left(\xi_{\omega\omega'}\right)=\mu_{\omega\omega'}a_{\omega'}-\nu_{\omega\omega'}a^{\dagger}_{\omega}\,,\qquad\text{and}\qquad\mathcal{S}^{\dagger}\left(\xi_{\omega\omega'}\right)a^{\dagger}_{\omega'}\mathcal{S}\left(\xi_{\omega\omega'}\right)=\mu_{\omega\omega'}a^{\dagger}_{\omega'}-\nu^*_{\omega\omega'}a_{\omega}\,,
\nonumber \\
\end{eqnarray}
and we have
 \begin{align}
    \langle \xi_{\omega\omega'}|a_{\omega}|\xi_{\omega\omega'}\rangle&=0\,,&\langle\xi_{\omega\omega'}| a_{\omega}\,a_{\omega'}|\xi_{\omega\omega'}\rangle&=-\mu_{\omega\omega'}\nu_{\omega\omega'}\,,&\langle\xi_{\omega\omega'}| a_{\omega}^{\dagger}a_{\omega'}|\xi_{\omega\omega'}\rangle&=\eta_{\omega\omega'}^2
2\pi \delta(\omega-\omega')\,, \end{align} where
$\mu_{\omega\omega'}=\cosh r_{\omega\omega'}$,
$\nu_{\omega\omega'}=e^{i\theta_{\omega\omega'}}\,\sinh
r_{\omega\omega'}$ and $\eta_{\omega\omega'}=|\nu_{\omega\omega'}|$.
Notice that the retarded Green's function defined in~(\ref{G_HR})
remains the same  in the two-mode squeezed vacuum state because
  the involved Bogoliubov transformations are the canonical
ones so they preserve the commutation relations between the creation
and annihilation operators. Then, the position correlator $\langle X(t) \, X(t') \rangle$
in the squeezed vacuum states can be calculated straightforwardly.  Using the
Langevin equation in~(\ref{langevin_eq}), which is also valid for a particle influenced by the environmental quantum fields in general quantum states,  we can find the corresponding  Hadamard function of the
boundary fields in the squeezed vacuum states  $G_{H}^{(s)}(t,t')$
\begin{align}\label{G_H_sv}
   { G_{H}^{(s)}(t,t') }
    &=\frac{\pi \alpha'}{2} \int_{0}^{\infty}\!\frac{d\omega}{2\pi}\int_{0}^{\infty}\!\frac{d\omega'}{2\pi}\; {W}(\omega) {W}(\omega') \, U_{\omega} \, U_{\omega'} \, G_R^{(0)} (\omega) \, G^{(0) *}_R (\omega') \notag \\
   &  \biggl[- \mu_{\omega\omega'} \nu_{\omega\omega'}\,{\frac{G_R^{(0)} (\omega')}{G_R^{(0) *} (\omega')}} e^{-i\omega t-i\omega' t'}\biggr.+\biggl.2\pi \delta(\omega-\omega')\Bigl( \eta_{\omega\omega'}^2+\frac{1}{2}
    \Bigr)\Bigl(e^{-i\omega t+i\omega'
    t'}\Bigr)\biggr]+\mbox{h.c.}\, ,
\end{align}
where $G_R^{(0)}$ is the Fourier transform of the retarded Green's
function  in the normal vacuum state obtained by~(\ref{G_R}). In
the above expression, we have introduced the simplest window
function ${W}(\omega)$  given by the unit-step functions
\begin{align}\label{w}
    {W}(\omega)&=1\,,&&\text{if $\omega_0-\Delta \leq\omega\leq\omega_0+\Delta$}\,.
\end{align}
Thus only modes within the frequency band $\omega_0 -\Delta
\leq\omega\leq\omega_0+\Delta$ are excited to the squeezed vacuum.
The other modes remain in normal vacuum. We can choose a more
smooth window function as long as it falls off to zero
sufficiently fast outside the frequency band of interest.
Apparently, in~(\ref{G_H_sv}) there are two distinct contributions
to the Green's function. The  second terms  of (\ref{G_H_sv}) are
the stationary component. However, there exists a nonstationary
component, and a time-dependent external field is required to
squeeze the vacuum. From the perturbed holographic influence
functional derived in~(\ref{S_phi}), the corrections to the
Hadamard function of boundary fields in vacuum~(\ref{G_0_H}),
denoted by $G_{H}^{(\phi)}(t,t')$, can be obtained as
\be\label{G_H_phi}
    G_{H}^{(\phi)}(t,t')= \int_{0}^{\infty}\!\frac{d\omega}{2\pi}\int_{0}^{\infty} \frac{d \omega'}{2\pi}\; 2 \, G_H^{(0)} (\omega)\,\left\{ \phi(\omega+\omega',r_b) \, e^{-i\omega t-i\omega'
    t'}+\phi(\omega-\omega',r_b) \, e^{-i\omega t+i\omega'
    t'}+\mbox{h.c.}\right\} \, .
  \ee
In the limits of small squeezing parameters and the narrow
bandwidth ( $\Delta/\omega_0 <1$ )  in (\ref{w}),  when $\omega$
and $ \omega'$ lie within the frequency band, we have $\omega
\approx \omega'$. Compared~(\ref{G_H_phi})  with~(\ref{G_H_sv}),
the field $\varphi(2 \omega)$ obtained from $\phi(\omega, r_b)$
in~(\ref{varphi}) can be related to the squeezing parameters up to
a constant phase by
\be \label{dual}
  {r_b^{-2-z}\varphi {(\omega+\omega' \approx 2
\omega)}= -{r_{\omega\omega}}  \, \Gamma (\frac{3}{2}+\frac{1}{z}
)\, \left(\frac{\omega}{z} \right)^{-\frac{2+z}{2z}}\, ,   }
  \ee
where the mode functions~(\ref{U}) can be expressed in terms of the
retarded Green's function and Hadamard function as

\be \label{U_G_fun}
 \frac{\pi^2 \alpha'}{2} \, U_{\omega}^2= \frac{G_H^{(0)} (\omega)}{ G_R^{(0)} (\omega) \, G_R^{(0) *} (\omega)} \, .
\ee
The large $r_b$ limit is taken in (\ref{dual}).
For a large but finite $r_b$, the squeezing parameters are typically small.

In the following we will consider the subvacuum effects for small squeezing
parameters so as to be consistent with the approximations we adopted
here. The above identification provides a possible scheme for the dual
gravity theory to generate metric perturbations that may correspond to the squeezed vacuum states of the boundary
fields, as suggested in~\cite{Lenny_99}. The
physical picture is that the gravitation waves are generated
somewhere at small $r$, and then propagate toward the boundary at
larger $r$. On their way to reach the boundary, the waves induce
fluctuations to the string dynamics, and thus excite the
quantum state of a string in such a way that their effects on the end point can be interpreted as those due to the squeezed vacuum fluctuations of the boundary field. However the faithful  identification of the dual
description needs the order by order comparison of the
correlators, obtained from  bulk theory and boundary field theory, in terms of small squeezing parameters. This work is
underway.

If we  choose the frequency-independent
squeezing parameters $\xi_{\omega, \omega'}=
2\pi \xi \delta(\omega-\omega')$  within the frequency band, the difference  between the velocity
dispersion due to squeezed vacuum and normal vacuum,  defined by $ \delta \langle (\Delta V)^2 \rangle_{\xi} \equiv
\langle (V (t) -V(0))^2\rangle_{\xi} - \langle (V(t)
-V(0))^2\rangle_{0}$ ,  can  be derived directly from the mode expansion~(\ref{force_mode_exp}),
\bea \delta \langle (\Delta V)^2 \rangle_{\xi}
&&=\int_{0}^{\infty}\!\frac{d\omega}{2\pi}\; {W}(\omega) \,  \frac{ 2 G_H^{(0)}
(\omega)}{G_R^{(0)} (\omega) G_R^{(0) *} (\omega)} {\omega^2} \biggl[{+} \mu_{\omega}
\nu_{\omega}\,(e^{-i\omega t}-1)^2 \biggr. \nonumber
\\
&& \quad \quad\biggl. {+} \mu_{\omega} \nu^{*}_{\omega} \,(e^{+i\omega
t}-1)^2\biggr.+\biggl. 2 \, \eta_{\omega}^2 (e^{-i\omega t}-1)
(e^{+i\omega t}-1)\biggr]\,\nonumber\\
&&= \int_{\omega_0-\Delta}^{\omega_0+\Delta}\!\frac{d\omega}{2\pi}\;
\frac{G_H^{(0)} (\omega)}{G_R^{(0)} (\omega) G_R^{(0) *} (\omega)} \, {\omega^2} 16 \,\big[
{-} \mu \, \eta \,\cos[\,\omega t\,-\theta]+\eta^2\big] \,
\sin^2\frac{\omega t}{2} \, , \label{delta_vf}
 \eea
where the mode functions can be written in terms of the retarded Green's function and Hadamard function~(\ref{U_G_fun}), and the window function~(\ref{w}) is introduced. Thus, the above expression of the shifted velocity has taken account of not only the stochastic effects of the environmental quantum fields through the Hadamard function, but also the dissipative effects of the retarded Green's function.
{The saturated value of the shifted velocity dispersion can be
found from the late-time behavior of (\ref{delta_vf}) in the limit $ (\omega_0
\pm \Delta) t \gg 1$. In this case the main contributions to the $\omega$-integration come from the regions of
small $\omega$.} The small $\omega$ expansion of $
G^{(0)}_H(\omega)/(G^{(0)}_R (\omega) G_R^{(0) *} (\omega))$ takes different forms
for $1<z<2$ and $ z>2$, and they are respectively given by
 \begin{align}
    \frac{G^{(0)}_H(\omega)}{G^{(0)}_R (\omega) G_R^{(0)*}(\omega)}&=
\frac{ 4\alpha'\, z
r_{b}^{z-2}}{\,\omega^{2}}\left[J^2_{\frac{1}{z}-\frac{1}{2}}\bigl(\frac{\omega}{z\,r_{b}^{z}}\bigr)+Y^{2}_{\frac{1}{z}-\frac{1}{2}}\bigl(\frac{\omega}{z\,r_{b}^{z}}\bigr)\right]^{-1}\,
\nonumber\\
& \simeq\begin{cases}
                                    \displaystyle \,\frac{2}{ \alpha'} \, \mathcal{N}_{1<z<2}\,m^{-2}\, \omega^{-3+2/z}\, ; \quad\quad   \mathcal{N}_{1<z<2}=\frac{1}{\Gamma^{2}(\frac{1}{z}+\frac{1}{2})},&1<z<2\,,\vspace{9pt}\\
                                    \displaystyle \, 2 \pi^2 \alpha' \,\mathcal{N}_{z>2}\, \omega^{-1-2/z}\,; \quad\quad \mathcal{N}_{z>2}=\frac{1}{ \Gamma^2(-\frac{1}{z}+\frac{1}{2})}\, ,
                                    &z>2\, .
                                \end{cases}
\end{align}
 There is a dramatic change at $z=2$. Notice that different
$\omega$ dependence in these two regimes of $z$ is mainly attributed to the fact that the low
frequency behavior of the retarded Green's function is dominated
by the inertial mass term when $1 < z < 2$ and by the
 $\gamma$ term when $z > 2$.

The change in the velocity dispersion at late time is given by
\begin{align}
 \delta \langle (\Delta V)^2 \rangle_{\xi}  & \simeq\begin{cases}
                                    \displaystyle \,\frac{4 \mathcal{N}_{1<z<2}}{ \pi \alpha'} \,{g_{+} (r,\theta)} \,\frac{[(\omega_0+\Delta)^{2/z}-(\omega_0-\Delta)^{2/z}]}{m^2}+\mathcal{O}((\omega_0\pm \Delta)^{-1+2/z}/ m^2 t)\,,1<z<2\, ;& \vspace{9pt}\\
                                    \displaystyle \,\frac{4 \pi  \alpha' \mathcal{N}_{z>2}}{ (z-1)} \,{g_{+} (r,\theta)} \,\bigg[  (\omega_0+\Delta)^{2-2/z}  -(\omega_0-\Delta)^{2-2/z} \bigg]+\mathcal{O}((\omega_0\pm
                                    \Delta)^{1-2/z}/t)\,,z>2\,,&
                                \end{cases}
\end{align}
where the function $g_{\pm}$ of squeezing parameters  is defined as \be
{g_{\pm}(\eta,\theta)=2\eta^2 {\pm} \eta \mu \cos(\theta)} \, .\label{gpm} \ee Thus, the
evolution of the shifted velocity dispersion of the particle due to the
squeezed  vacuum of environment fields will reach a saturated value at
late times, following a power-law like $t^{-1}$.

Notice that for some particular choices of squeezing parameters, the
function $g(\eta,\theta)$ can be negative, leading to the so-called
subvacuum phenomenon. The most negative value can be found as \be
\label{g_function}\eta^2-\frac{1}{2} \eta \mu \ge
-\frac{2-\sqrt{3}}{4} > -\frac{1}{2} \, . \ee
Therefore, the subvacuum effect has a lower bound given by the
inequality above. This is expected because the sum of the
velocity dispersion arising from the normal vacuum and the shifted
value due to the squeezed vacuum must be positive. Thus, the
shifted value has to be greater than the minus of the result from
the pure vacuum, which in turn constrains how negative the shifted value can reach.

In the
case of $1<z<2$, the subvacuum phenomenon shown in the velocity dispersion is found to have
the $1/m^2$ dependence, and is consistent with the
findings in weakly coupled fields~\cite{Lee_12} for the case of $z=1$. Although this
subvacuum effect is enhanced by a strongly coupling constant
$\lambda \propto 1/ \alpha'$, the heavy mass dependence $m^2
\propto \lambda^2$ will suppress this effect. As to the $z>2$
case, since the dominant term of the retarded Green's function
comes from the $\gamma$ term in~(\ref{G_R_LowF}) and
(\ref{gamma}), the coupling constant dependence in the $\gamma$
term leads to the $1/\lambda$ dependence of the subvacuum effect, which
is also relatively weak for a large $\lambda$. To
make an estimate, the typical length scale $L$, which is associated with the
breakdown of Lorentz invariance in quantum critical  theory,  is
introduced~\cite{Visser_09}.  If the scale $1/L$ is the largest
momentum scale in the system, the large mass $m$ can be
parameterized as $ m=\lambda/L$. Then, we find that
\begin{align}
 \vert \delta \langle (\Delta V)^2 \rangle_{\xi} \vert  & \simeq\begin{cases}
                                    \displaystyle \,\frac{1}{\lambda}  \bigg( \frac{L}{\lambda_0} \bigg)^{2/z}  \bigg( \frac{\Delta}{\omega_0} \bigg)\,, \quad\quad 1<z<2\, ;& \vspace{9pt}\\
                                    \displaystyle \,\frac{1}{\lambda} \bigg( \frac{L}{\lambda_0} \bigg)^{2-2/z}  \bigg( \frac{\Delta}{\omega_0} \bigg)\,, \quad\quad z>2\,,&
                                \end{cases}
\end{align}
where $\lambda_0$ is a typical wavelength of squeezed vacuum
modes, and $\lambda_0 > L$ in general. The result of the shifted
velocity dispersion is suppressed by a large $\lambda$, but in
principle observable.

\section{decoherence and recoherence}\label{sec3}
The above influence functional can also be applied to study the
nature of quantum coherence  of a particle when it
interacts with  the environment. We consider the initial state
$\bigl|\Psi(t_i)\bigr>$ of the particle to be a coherent
superposition of two localized states.  Additionally,  both states
can be arranged to have the same spatial point at the moment
$t_i$,
\begin{equation}
    \bigl|\Psi(t_i)\bigr>=\bigl|\psi_1(t_i)\bigr>+\bigl|\psi_2(t_i)\bigr>\,,
\end{equation}
then the initial density matrix of the state can be written as
\begin{eqnarray}
    \rho_q(t_i)=\bigl|\Psi(t_i)\bigr>\bigl<\Psi(t_i)\bigr|=\rho_{11}(t_i)+\rho_{22}(t_i)+\rho_{21}(t_i)+\rho_{12}(t_i)\,,
\end{eqnarray}
where
$\rho_{mn}(t_i)=\bigl|\psi_m(t_i)\bigr>\bigl<\psi_n(t_i)\bigr|$.
We assume that these two localized states $\bigl|\psi_m\bigr>$
move along their respective paths $C_{m}$ {such
that} they leave from the same spatial point and recombine at the
location ${q}_f$ at later time $t_f$. The interference pattern of
the superposed state at $t_{f}$ is given by the off-diagonal elements
of the reduced density matrix. Here we assume that the de Broglie
wavelength of the localized states is much shorter than the
 length scale of interest
so that the width of the wavefunctions  and their subsequent
 spreading can be legitimately neglected~\cite{HS}.
Thus, the leading effect of the reduced density matrix can be
obtained by evaluating the propagating function \eqref{propagator}
along a mean trajectory of the localized states dictated by an
external force. Then $\rho_r({q}_f,{q}_f,t_f)$ now becomes
\begin{equation}
    \rho_r({q}_f,{q}_f,t_f)=\bigl|\psi_1({q}_f,t_f)\bigr|^2+\bigl|\psi_2({q}_f,t_f)\bigr|^2+2\,e^{\mathcal{W}[\,\bar{q}^+,\bar{q}^-]}\;\operatorname{Re}\left\{e^{i\,\Phi[\,\bar{q}^+,\bar{q}^-]}\psi_1^{\vphantom{*}}({q}_f,t_f)\,\psi_{2}^{*}({q}_f,t_f)\right\}\,,
\end{equation}
where the $\mathcal{W}$ and $\Phi$ functionals are the phase and
modulus of the influence functional defined by:
\begin{equation}
    \mathcal{F}[\,q^+,q^-]=\exp\Big\{\mathcal{W}[\,q^+,q^-]+i\,\Phi[\,q^+,q^-]\Bigr\}\,,
\end{equation}
and they are
\begin{align}
    \Phi[\,{\bar q}^+,{\bar q}^-]&=-\frac{1}{2}\int\!dt \!\!\int\!dt'\Bigl[\,\bar{q}^+(t)-\bar{q}^-(t')\Bigr] {G_{R}(t,t')}\Bigl[\,\bar{q}^+(t) + \bar{q}^-(t')\Bigr]\,,\label{phase}\\
    \mathcal{W}[\,{\bar q}^+,{\bar q}^-]&=-\frac{1}{2}\int\!dt \!\!\int\!dt'\Bigl[\,\bar{q}^+(t)-\bar{q}^-(t')\Bigr] {G_{H}(t,t')}\Bigl[\,\bar{q}^+(t) - \bar{q}^-(t')\Bigr]\,,\label{decoherence}
\end{align}
 evaluated along the classical trajectories, ${C}_1=\bar{q}^{+}$
and ${C}_2=\bar{q}^{-}$. The modulus of the influence functional
$\mathcal{W}$ reveals decoherence between the coherently superposed states, and the
phase functional ${\Phi}$ results in an overall phase shift for the
interference pattern. Both effects to the particle states arise from
the interaction with quantum fields. The retarded Green's function
and Hadamard function constructed out of quantum critical fields
have been obtained by the holographic method.

In~\cite{Lee_06, Yeh_14}, we studied the decoherence effect to a
quantum particle from {the vacuum state} of electromagnetic fields
and quantum critical fields respectively. However from what we
have learned above, the retarded Green's function of strongly
coupled quantum critical fields plays an essential role in determining the trajectory, otherwise, driven by an
external or noise force. In addition, the terms that dominate in the regime $1<z<2$
and $z>2$ give rise to rather different relaxation behaviors of
the particle.
 Accordingly, we will reexamine this effect by taking account of the retarded
Green's function properly into the equation of motion.  So, for a
prescribed force whose Fourier transform is defined by
$F_{ex}(\omega) \equiv  m \omega^2 {\zeta (\omega)}$, the trajectory
follows the solution of~(\ref{Langevin_sol}) is, \be q_{\zeta}
(\omega)= m \, \omega^2 \, \zeta(\omega) /G_R (\omega) \, , \ee
where the form of $\zeta$ will be specified later. We consider the mean trajectories of two localized states specified by
$\bar{q}^{\pm} =\pm q_{\zeta}$ in the interference experiment and
then for the time-translation invariant Green's function, the
decoherence functional is given by
 \be \label{W}
{\mathcal{W}}=-2 \int \frac{d\omega}{2\pi}  \, \frac{ m \,
 \omega^2 \, \zeta (\omega)}{ G_R (\omega)} \, G_H (\omega) \, \frac{
 m \,
 \omega^2 \, \zeta (-\omega)}{ G^*_R (\omega)}  \, .\ee The function $\zeta(t)$
is required to be sufficiently smooth and is chosen to take the
form~\cite{HS},
\begin{equation}\label{path}
    \zeta(t)=\frac{\ell_0}{\tau_0^4} \, (t^2-\tau_0^2 )^2\,,
\end{equation}
where $2\ell_0$ characterizes a length scale for path separation
and $2\tau_0$ is the effective flight time for $-\tau_0 < t < \tau_0$ as in \cite{Lee_06}. These two prescribed
classical trajectories are symmetric with respect to the initial
position so that $q^{+}(t)+q^{-}(t)=0$ and thus we will not see
any phase shift, $\Phi =0$ from~(\ref{phase}). We now focus on the
decoherence functional $\mathcal{W}$ contributed by the quantum
critical fields.
 For the environment in its normal vacuum state,  the retarded Green's function $G_R^{(0)}$ and  Hadamard function
  $G_H^{(0)}$ are given in~(\ref{G_R}) and (\ref{G_0_H}) respectively, and then the decoherence
 functional is found to be
\begin{align}
  \mathcal{W}_0  & \simeq\begin{cases}
                                    \displaystyle \, -\frac{2 \mathcal{N}^{\mathcal{W}_0}_{1<z<2}}{ \pi \alpha'}  \, \frac{\ell_0^2}{\tau_0^{2/z}}+ {\mathcal
O}(1/r_b^2\tau_0^2)\, , \quad\quad 1<z<2 \, ;& \vspace{9pt}\\
                                    \displaystyle \,-2 \pi \,  \alpha'\,  \mathcal{N}^{\mathcal{W}_0}_{z>2} \,  \frac{ m^2 \,\ell_0^2}{\tau_0^{2-2/z}} + {\mathcal
O}(1/r_b^2\tau_0^2) \,, \quad\quad z>2 \, ,&
                                \end{cases}
\end{align}
where \bea \mathcal{N}^{\mathcal{W}_0}_{1<z<2} &=& 4
\mathcal{N}_{1<z<2} \frac{2^{12-\frac{2}{z}}}{z^2(4z-1)} (5z-2)(z-1)
\, \cos
(\frac{\pi}{z})\, \Gamma (-6+\frac{2}{z})\, ,\nonumber \\
 \mathcal{N}^{\mathcal{W}_0}_{z>2} &=&  4 \mathcal{N}_{z>2} \frac{ 2^{10+\frac{2}{z}}}{z^2(3z+1)}\, (3z+2) (z-1) \, \cos(\frac{\pi}{z}) \, \Gamma (-4-\frac{2}{z} ) \nonumber \, .
 \eea
The negative value of the $\mathcal{W}$ function indicates the decoherence effect on the wavefunction of the particle.
 Thus, in the case $1<z<2$, for
a fixed travel time $\tau_0$, the magnitude of $\vert \mathcal{W}_0 \vert$  increases as $z$
increases, and a large coupling constant $\lambda \propto 1/\pi
\alpha'$ renders significant decoherence. As for $z>2$, the combined
results from the respective $1/\lambda$ and $m^2 \propto \lambda^2 $
dependence  also cause large decoherence. In this range of $z$, the value of
  $\vert \mathcal{W}_0 \vert$ decreases as $z$ increases instead.
The most significant decoherence effect occurs at $z=2$.  For $z=1$,
the result is consistent with the  weakly
coupled relativistic field case apart from its large coupling constant
dependence, giving large decoherence.

Now we study the $\mathcal{W}$ function associated with the squeezed
vacuum state.  The above expression~(\ref{W}) needs to be modified to
account for the non-stationary component of the Green's function.
The associated decoherence functional is defined by
$\mathcal{W}_{\xi} =\mathcal{W}_0 + \delta \mathcal{W}_{\xi}$ where
$\delta \mathcal{W}_{\xi}$ is the modification from the result  due
to the normal vacuum  of the field, and is given by \be
\label{delta_W} \delta \langle \mathcal{W} \rangle_{\xi} =-16
\int_{\omega_0-\Delta}^{\omega_0+\Delta}\!\frac{d\omega}{2\pi}\;
\frac{G_H^{(0)} (\omega)}{G_R^{(0)} (\omega) G_R^{(0) *} (\omega)} \, m^2
\omega^4 \, \zeta(\omega) \, \zeta (-\omega) \, \big[ \mu \, \eta
\,\cos[\,\omega \, t\,-\theta]+\eta^2\big]  \, .
 \ee
The straightforward calculations show at large time scales $ \tau_0
\gg 1/\omega_0 $,
 \begin{align}\label{W_xi}
  \delta \mathcal{W}_{\xi}  & \simeq\begin{cases}
                                    \displaystyle \, - \frac{2 \mathcal{N}_{1<z<2}}{\pi \alpha'} \, {g_{-}(r,\theta)}\,\frac{\ell_0^2}{\tau_0^{2/z}} \,\frac{1024}{(\omega_0 \tau_0)^{4-2/z}}\, \frac{\Delta}{\omega_0} \bigg[ \sin(\omega_0 \tau_0) + \mathcal{O}( 1/(\omega_0\tau_0) \bigg] \,,\,\,  1<z<2\, ;& \vspace{9pt}\\
                                    \displaystyle \,-2  \pi \, \alpha'\,  \mathcal{N}_{z>2} \,{g_{-}(r,\theta)}\,  m^2\, \frac{\ell_0^2}{\tau_0^{2-2/z}}   \,\,\frac{1024}{(\omega_0 \tau_0)^{2+2/z}}\, \frac{\Delta}{\omega_0} \bigg[ \sin(\omega_0 \tau_0) + \mathcal{O}( 1/(\omega_0\tau_0) \bigg] , \,\,\, z>2 \, ,&
                                \end{cases}
\end{align}
where we have assumed the narrow bandwidth, namely $\omega_0 >
\Delta$.
Thus, by choosing some particular values of squeezing parameters,
$g_{-}(r,\theta)$ {in~(\ref{gpm})}  can be negative  with the most negative value $-(2-\sqrt{3})/2$.
Thus, an
increase of contrast, as compare to that from the normal
vacuum, is seen for $ \delta \mathcal{W}_{\xi} >0$ due to $g_{-}(r,\theta)<0$ for certain values of the squeezing parameters.  This is the phenomenon of recoherence, which is another
subvacuum effect. For $1<z<2$, a strong coupling $\lambda \propto
1/ \alpha'$  may
render an enhancement on this recoherence effect. For $z>2$,
this subvacuum effect is also enhanced due to large $m^2/\lambda
\propto \lambda$ dependence.
To have a rough
estimate, we consider the typical frequency of the squeezed vacuum
modes  such that $(1/\omega_0)\equiv
\lambda_0 \sim \ell_0$~\cite{HS}. When $g_{-}(r,\theta)<0$ in the expression~(\ref{W_xi}) for appropriate squeezing parameters, an order of the
magnitude estimation of the recoherence phenomenon  is found  to be
 \begin{align}
  \delta \mathcal{W}_{\xi}   & \simeq\begin{cases}
                                    \displaystyle \,  \lambda \,\bigg(\frac{\ell_0}{\tau_0}\bigg)^4 \,\bigg(\frac{\ell_0}{L}\bigg)^{2-2/z}\, \bigg(\frac{\Delta}{\omega_0}\bigg)\,,\quad \quad  1<z<2 \, ;& \vspace{9pt}\\
                                    \displaystyle \, \lambda \,\bigg(\frac{\ell_0}{\tau_0}\bigg)^4 \,\bigg(\frac{\ell_0}{L}\bigg)^{2/z}\, \bigg( \frac{\Delta}{\omega_0} \bigg) \,, \quad \quad z>2 \, ,&
                                \end{cases}
\end{align}
where we have again parameterized $m \simeq \lambda/L$, and $L$, the Lorentz
symmetry breaking scale in quantum critical field theory, is introduced. In the
$z=1$ case, the
$(\ell_0/\tau_0)^4$ dependence
is consistent with the findings from an environment of a weakly coupled relativistic field~\cite{HS}.  For a general $z$, this phenomenon will also
depend on $\ell_0/L$. When $\ell_0
> L$, the maximum recoherence effect is found at $z=2$ whereas the
minimum effect happens at $z=1$ and $z \rightarrow \infty$. The
presence of vacuum fluctuations of strongly coupled quantum
critical fields in squeezed state gives rise to
potentially large recoherence effect on the
wavefunction of a particle. Thus the large coupling constant may
offer the possibility to observe these subvacuum phenomena.

\section{Summary and outlook}\label{sec4}
Using the holographic influence functional approach, the effects of the strongly coupled quantum critical fields
on the dynamics of a  particle are studied. The dual description is a string moving in
4+1-dimensional Lifshitz geometry. We first study the influence on
the velocity dispersion of a  particle  from the squeezed
vacuum states of the quantum critical fields. We find that the
evolution of the velocity dispersion  will
reach a constant at late times where its relaxation dynamics
follows a power law in time as $t^{-1}$. With particular choices
of squeezing parameters, the saturated value
 is found to be smaller than the value given by the normal
vacuum background. This leads to a subvacuum effect. The reduction
in the velocity dispersion changes dramatically as the dynamical
exponent passes through $z = 2$. This subvacuum effect, which is
found proportional to $1/\lambda \propto \alpha' $, is suppressed
by a large value $\lambda$ in quantum field theory, but is in
principle observable. We then study the decoherence dynamics of a
quantum particle induced also by the squeezed vacuum fluctuations. We
find that coherence loss can be shown less severe in certain squeezed vacuums than in normal vacuum.
This recovery of coherence is understood as recoherence, another manifestation of the subvacuum phenomena.
We make some estimates of the degree of recoherence,
which is enhanced by a large coupling constant $\lambda$, and thus can
potentially be measurable. We also show that there exists a bound to constrain the above-mentioned
subvacuum phenomena.

Finally,  we would like to  point out some of our future works. In
view of a close relation between the holographic approach and the
field-theoretical study, an
important next step is to study the issue of negative energy
density and find the associated quantum inequality, which
constrains the magnitude and duration of the negative energy density in the strongly coupled
field, using the method of the holography principle. This might give a profound
implication to the existence of exotic spacetimes sourced by
negative energy.

\begin{acknowledgments}
We would like to thank Jen-Tsung Hsiang for collaboration on the
early stage of this work. This work was supported in part by the
Ministry of Science and Technology, Taiwan.
\end{acknowledgments}

\end{document}